\documentclass[11pt]{article}
\usepackage{geometry}
\usepackage{graphicx}
\usepackage{tensor}
\usepackage{url}
\usepackage{bm}
\usepackage{bbm}
\usepackage{amsthm}
\usepackage{amsfonts}
\usepackage{tabto}
\usepackage[english]{babel}
\usepackage[T1]{fontenc}
\usepackage[utf8]{inputenc}
\usepackage{xcolor}
\usepackage{tensor}
\usepackage{commath}
\usepackage{mathtools}
\usepackage{braket}
\usepackage{physics}
\usepackage{float}
\usepackage{subfig}
\usepackage[hidelinks]{hyperref}
\usepackage{cleveref}
\usepackage{authblk}
\usepackage{amsmath}
\usepackage{amssymb}
\usepackage{tabu}

\usepackage{soul}
\usepackage{cancel}

\hypersetup{
colorlinks=true,
linkcolor=blue,
citecolor=red,
urlcolor=cyan
}

\newcommand{\ebd}{\vcentcolon =}

\newcommand{\intt}[1]{\int\dd[3]{#1}}
\newcommand{\inttt}[1]{\int\dd{#1}}

\makeatletter
\newcommand{\fmarki}{*}
\newcommand{\fmarkii}{\ensuremath{\dagger}}
\newcommand{\fmarkiii}{\ensuremath{\ddagger}}

\def\@fnsymbol#1{{\ifcase#1\or \fmarki\or \fmarkii\or \fmarkiii \else\@ctrerr\fi}}
\makeatother

\renewcommand{\fmarki}{$\natural$}
\renewcommand{\fmarkii}{$\star$}
\renewcommand{\fmarkiii}{$\mathsection$}

\begin{document}

\title{{\bf Constraints on quantum spacetime-induced  decoherence from neutrino oscillations}}
\date{}
\author[]{
Vittorio D'Esposito\thanks{vittorio.desposito@unina.it} \and Giulia Gubitosi\thanks{giulia.gubitosi@unina.it}}
\affil[]{Dipartimento di Fisica “E. Pancini”, Università di Napoli Federico II \\ INFN sezione di Napoli,
Complesso Universitario di Monte S. Angelo Edificio 6, via Cintia, 80126 Napoli, Italy}

\maketitle

\begin{abstract}
We investigate the implications of decoherence induced by quantum spacetime properties on neutrino oscillation phenomena.
We develop a general formalism where the evolution of neutrinos is governed by a Lindblad-type equation and we compute the oscillation damping factor for various models that have been proposed in the literature. 
Furthermore, we discuss the sensitivity to these effects of different types of neutrino oscillation experiments, encompassing astrophysical, atmospheric, solar, and reactor neutrino experiments. By using neutrino oscillation data from long-baseline reactors and atmospheric neutrino observations, we establish stringent constraints on the energy scale governing the strength of the decoherence induced by stochastic metric fluctuations, amounting to, respectively, $E_{QG}\geq 2.6 \cdot 10^{34}\; \text{GeV}$ and $E_{QG}\geq 2.5\cdot 10^{55}\;\text{GeV}$.

\end{abstract}

\section{Introduction}

Propagation in a quantum spacetime can affect the pattern of neutrino oscillations. Specifically, this may be ascribed to deformations of the neutrino dispersion relation \cite{Christian:2004xb}, minimal length and generalized uncertainty principle scenarios \cite{Sprenger:2011jc, Banerjee:2022slh}, $CPT$ violations \cite{Mavromatos:2007hv, Mavromatos:2004sz, Barenboim:2004wu}, Lorentz invariance violations \cite{IceCube:2017qyp,IceCube:2021tdn} or generalized quantum mechanical evolution equations leading to decoherence mechanisms and collapse models \cite{Christian:2005qa, ELLIS199237, Ellis:1997jw}. Some of these models cause deviations from the standard neutrino oscillation pattern, while others \cite{Mavromatos:2007hv, IceCube:2017qyp, IceCube:2021tdn} imply a damping of the total neutrino fluxes (the two possibilities are discussed and compared in \cite{Blennow:2005yk}). 

In this work, we investigate how neutrino oscillations are affected by fundamental decoherence, a generic feature emerging in quantum gravity research \cite{Arzano:2022nlo,Petruzziello:2020wkd,Breuer:2008rh, Diosi1984, Percival:1995an, Gambini:2004de, Bassi:2017szd}. To account for decoherence, the usual the Schr\"{o}dinger equation describing propagation of neutrino wavepackets is replaced by a general Lindblad equation \cite{Lindblad1976}. 
We fully develop the relevant formalism, and show that a Lindblad-type evolution introduces an exponential damping factor into the oscillation probability that quenches the oscillations, without altering the neutrino fluxes. The functional dependence of the damping factor on the neutrino energy characterizing different neutrino experiments and on neutrino masses depends on the specific decoherence model. We compute it explicitly for a number of quantum spacetime-induced decoherence mechanisms originating from different sources: non-commutative spacetime \cite{Arzano:2022nlo}, stochastic fluctuations of a minimal length \cite{Petruzziello:2020wkd}, stochastic fluctuations of metric perturbations around Minkowski spacetime \cite{Breuer:2008rh}, and the interaction with a thermal background of gravitons \cite{Blencowe:2012mp}. We discuss the relevance of the resulting oscillation damping for measurements exploiting reactor, solar, atmospheric and astrophysical neutrino. 

Because the experiment baseline generally enters linearly in the oscillation damping factor one could naively assume that long-baseline experiments might provide stronger constraints on the effect. However, for very long baselines neutrino wavepackets decohere already in the standard oscillation scenario, because of the different propagation velocities of the mass eigenstates. Therefore, oscillations are washed out and flavour eigenstates decohere into mass eigenstates \cite{Farzan:2008eg,Akhmedov:2009rb, Addazi2022, Giunti:1997wq}, thus rendering the fundamental decoherence effect irrelevant. 

We find that the most sensitive probes for fundamental decoherence are reactor and atmospheric neutrinos. Their sensitivity to decoherence depends on the dependence of the effect on the neutrino energy and masses.
Among the models we considered, we find that the only one that can be significantly constrained using these observations is the one where decoherence is induced by metric fluctuations \cite{Breuer:2008rh}. For this model we establish stringent constraints on the quantum gravity scale governing the strength of the effect. Even though this is the only model that can be meaningfully constrained using neutrino oscillations,  for completeness  we discuss the dependence on the relevant experimental parameters of all the  models we considered, to show in detail why neutrino oscillations do not offer a good phenomenological window for them.

The plan of the paper is as follows. In \cref{standard oscillations section} we review the standard derivation of neutrino oscillation probability in vacuo, including a discussion on standard decoherence effects. In \cref{decoherence sec} we show how to adapt the standard formalism to account for fundamental decoherence and derive the general transition probability. We specialize our computations to several decoherence models, computing explicitly the resulting oscillation damping factors in a two-flavour setting. In \cref{sec:experiments}
we discuss the sensitivity to fundamental decoherence of different classes of neutrino oscillation experiments, including astrophysical, atmospheric, solar and reactor neutrinos, showing that only atmospheric and reactor neutrino experiments are potentially relevant. We derive the neutrino energy range that would make the various decoherence models relevant for these kind of observations. Finally we establish constraints on decoherence induced by stochastic metric fluctuations by using long-baseline reactor neutrino data from the KamLAND experiment and atmospheric neutrino data from the Super-Kamiokande experiment.

\section{Neutrino oscillation in the standard scenario}\label{standard oscillations section}

Neutrino oscillations are commonly described within the plane wave approximation \cite{Akhmedov:2009rb}, which requires specific assumptions concerning the conditions at production (equal energy, equal momenta, equal velocity or energy-momentum conservation). This is reasonable in the standard neutrino oscillation picture, and all of the possible assumptions produce the correct oscillation probability. However, it is not suited for decoherence analyses, since typically the decoherence terms in the evolution break the equivalence of the assumptions, resulting in ambiguous results. For this reason, in the following, we describe neutrino evolution within the wave packet formalism. We start by briefly reviewing the derivation in the standard scenario following \cite{Akhmedov:2009rb}, and then generalize the analysis by introducing decoherence effects. We remark that a fully consistent treatment of neutrino oscillations should require a relativistic quantum field theory framework. However, the correct (standard) oscillation probability can also be derived with a first quantization approach, see, \emph{e.g.}, \cite{Akhmedov:2009rb}. It is then conceivable to discuss possible decoherence effects affecting the standard oscillations pattern within the same framework, as extensively done in the literature \cite{Christian:2004xb,Banerjee:2022slh,Mavromatos:2007hv,Mavromatos:2004sz,Barenboim:2004wu,Christian:2005qa,Giunti:1997wq,DeRomeri:2023dht,Lisi:2000zt,Adler:2000vfa}.

The wave packet corresponding to a generic neutrino flavour state $\ket{\nu_\gamma}$ can be written as a superposition of mass states wave packets $\ket{\psi_i} \otimes \ket{\nu_i}$ as 
\begin{equation}
    \ket{\nu_\gamma} = \sum_i U^*_{\gamma i} \ket{\psi_i}\otimes\ket{\nu_i} = \sum_i U^*_{\gamma i} \intt{p}\psi_i\qty(\pmb{p})\ket{\pmb{p}}\otimes\ket{\nu_i}\;,\label{Neutrino state}
\end{equation}
where $U$ is the neutrino mixing matrix and $\ket{\pmb{p}}$ are momentum eigenstates. If the neutrino state is described by a density operator $\rho(t)$ at time $t$, such that $\rho(0) = \ketbra{\nu_\beta}{\nu_\beta}$, the transition probability from the initial flavour state $\beta$ to a different flavour state $\alpha$ within time $t$ is given by 
\begin{equation}
    P(\beta\to\alpha;t) = \Tr{\rho(t)\ketbra{\nu_\alpha}{\nu_\alpha}}\;.
\end{equation}
Taking into account the time evolution of the density matrix
\begin{equation}
    \rho(t) = e^{-iHt} \qty[\ketbra{\nu_\beta}{\nu_\beta}] =
    \sum_{i,j} U_{\beta i}^* U_{\beta j} \int \text{d}^3p \,\text{d}^3 q\,\psi_i\qty(\pmb{p})\psi_j^*\qty(\pmb{q}) e^{-i\qty[E_i\qty(\pmb{p})-E_j\qty(\pmb{q})]t}\ketbra{\pmb{p}}{\pmb{q}}\otimes \ketbra{\nu_i}{\nu_j}\label{time evolution standard}
\end{equation}
and decomposing the flavour states as in \eqref{Neutrino state},

\begin{equation}
    \ketbra{\nu_\alpha}{\nu_\alpha} = \sum_{k,l} U_{\alpha k}^*U_{\alpha l} \ketbra{\phi_k}{\phi_l}\otimes \ketbra{\nu_k}{\nu_l}\;,
\end{equation}
the transition probability at time $t$ reads
\begin{equation}\label{oscillation probability standard wave-packet}
     P(\beta\to\alpha;t) = \sum_{i,j} U^*_{\beta i} U_{\beta j} U^*_{\alpha j} U_{\alpha i} \int \text{d}^3p \,\text{d}^3 q\,\psi_i\qty(\pmb{p})\psi_j^*\qty(\pmb{q})\phi_j\qty(\pmb{q})\phi_i^*\qty(\pmb{p})e^{-i\qty[E_i\qty(\pmb{p})-E_j\qty(\pmb{q})]t}\;,
\end{equation}
where we used the relation
\begin{equation}
    \Tr{\qty(\ketbra{\pmb{p}}{\pmb{q}}\otimes \ketbra{\nu_i}{\nu_j})\Big(\ketbra{\phi_k}{\phi_l}\otimes \ketbra{\nu_k}{\nu_l}\Big)} = \phi_k\qty(\pmb{q})\phi_l^*\qty(\pmb{p})\delta_{kj}\delta_{li}\,.
\end{equation}
Note that the plane-wave description can be recovered from \eqref{oscillation probability standard wave-packet} by setting $\phi_i(\pmb{p}) = e^{i\pmb{p}\cdot\pmb{x}}$, $\psi_i(\pmb{p}) = \delta^{(3)}(\pmb{p}-\pmb{p}_i)$ and performing one of the standard space-to-time conversion techniques \cite{Akhmedov:2009rb}. 

 Since in any neutrino oscillation experiment the propagation distance is much larger that the wave-packet size, the problem is effectively one-dimensional, and only the component of momenta in the direction between the source and the detector is relevant. If the wave-packets of the produced and detected neutrinos are peaked, respectively, in $x=0$ and $x=L$, the corresponding momentum distribution functions are $\psi_i(p)=f_i^{(S)}(p-p_i)$ and $\phi_i(p) = f_i^{(D)}(p-p^\prime_i)e^{-ipL}$. Assuming that the momentum distribution functions are sharply peaked around the mean momenta $p_i$ (at emission) and $p_i^\prime$ (at detection), one can expand the energies $E_{i,j}(p)$ about the mean momenta at first order:
\begin{equation}
    E_{i,j}(p)
    \simeq E_{i,j}(p_{i,j}) + (p-p_{i,j})v_{g_{i,j}}\;,
\end{equation}
where $v_{g_{i,j}}=p_{i,j}/E_{i,j}(p_{i,j})$ is the group velocity of the wave-packet. Then the transition probability \eqref{oscillation probability standard wave-packet} reads

\begin{eqnarray}
    &&P(\beta\to\alpha;t) = \sum_{i,j} U^*_{\beta i} U_{\beta j} U^*_{\alpha j} U_{\alpha i} \,e^{-i\Delta E_{ij}t+i\Delta p_{ij}L} \inttt{p}\dd{q}\,f_i^{(S)}(p){f_j^{(S)}}^*(q) \nonumber \\ &&\cdot f_j^{(D)}(q+\delta_j) e^{-iq L} {f_i^{(D)}}^*(p+\delta_i) e^{ipL} e^{-i\qty[pv_{g_i}-qv_{g_j}]t}\;,\label{prob}
\end{eqnarray}
where we further shifted the integration variables $p-p_i \to p$ and $q-p_j \to q$ and introduced the notation $\delta_i = p_i-p_i^\prime$, $\Delta E_{ij} = E_i(p_i)-E_j(p_j)$ and $\Delta p_{ij} = p_i-p_j$.

Because the emission and arrival times of neutrinos are not measured in typical oscillation experiments, the transition probability \eqref{prob} is averaged over time \cite{Akhmedov:2009rb} and a normalization factor is introduced so that the probability is normalized, $\sum_{\alpha,\beta} P(\beta\to \alpha) = 1$:
\begin{eqnarray}\label{probbb}
    &&P(\beta\to\alpha) = \mathcal{N}\sum_{i,j} U^*_{\beta i} U_{\beta j} U^*_{\alpha j} U_{\alpha i} \,e^{i\Delta p_{ij}L} \inttt{t}\inttt{p}\dd{q}\,f_i^{(S)}(p){f_j^{(S)}}^*(q) \nonumber \\ &&\cdot f_j^{(D)}(q+\delta_j) {f_i^{(D)}}^*(p+\delta_i) e^{i(p-q)L} e^{-i\qty[pv_{g_i}-qv_{g_j}+\Delta E_{ij}]t}\;,
\end{eqnarray}
where $\mathcal{N}$ is the normalization factor. Integration over the time variable results in a $\delta(pv_{g_i}-qv_{g_j}+\Delta E_{ij})$. This is simply enforcing an equal-energy condition on the wave packet components, $\delta\qty(E_i(p)-E_j(q))$. This condition arises because only the equal-energy components are not washed out by the rapidly oscillating factor in the integration. Therefore, the integration over time can be replaced by enforcing the equal-energy condition on the wave packets components. This produces equivalent results as to the time integration when no decoherence is in place (the complete proof can be found in \cite{Akhmedov:2009rb} and will also be discussed in \cite{PhDVittorio}), and it is the only one that gives well-defined results once decoherence processes are accounted for, as we show in the following section. 

Calling $r_{ij}=v_{g_i}/v_{g_j}$, one finally obtains
\begin{eqnarray}
    &&P(\beta\to\alpha) = \sum_{i,j} U^*_{\beta i} U_{\beta j} U^*_{\alpha j} U_{\alpha i} \,e^{i \phi_{ij}} \frac{\mathcal{N}}{v_{g_j}}\inttt{p} f_i^{(S)}(p) {f_j^{(S)}}^*(r_{ij}\,p+\Delta E_{ij}/v_{g_j}) \nonumber \\ &&\cdot f_j^{(D)}(r_{ij}\,p+\Delta E_{ij}/v_{g_j}+\delta_j) {f_i^{(D)}}^*(p+\delta_i) e^{ip(1-r_{ij})L}\;,\label{oscillation probability standard wave-packet final}
\end{eqnarray}
where
\begin{equation}
    \phi_{ij} \ebd \qty(\Delta p_{ij}-\frac{\Delta E_{ij}}{v_{g_j}})L \simeq  \qty(\Delta p_{ij}-\frac{\Delta E_{ij}}{v_{g_{ij}}})L\;,
\end{equation}
is the phase factor and the last approximation, in which $v_{g_{ij}} \ebd (v_{g_i}+v_{g_j})/2$, can be performed because the difference between $v_{g_i}$ and $v_{g_{ij}}$ is of order $\Delta m^2_{ij}$ and $\Delta E_{ik}$ is already of order $\Delta m^2_{ij}$. For relativistic or quasi-degenerate neutrinos, the momentum difference can be expanded in terms of the the energy difference and the mass difference $\Delta m^2_{ij}$
\begin{equation}
    \Delta p_{ij} \simeq \frac{\Delta E_{ij}}{v_{g_{ij}}} - \frac{\Delta m^2_{ij}}{2\,p_{ij}}\;,
\end{equation}
where $p_{ij}$ is the average momentum. This yields the standard phase

\begin{equation}\label{phase factor}
     \phi_{ij} = - L\frac{\Delta m^2_{ij}}{2\,p_{ij}}\;.
\end{equation}

In general, the last phase factor in eq. \eqref{oscillation probability standard wave-packet final} can be rapidly oscillating, and this would produce a loss of coherence for the neutrino wave packets\footnote{For other possible causes of decoherence see the discussion in \cite{Ohlsson:2000mj}.}. However, in several experimental setups of interest, this effect is negligible. This is the case when two conditions are met. 

The \emph{propagation coherence condition} ensures that the distance $L$ traveled by the neutrinos is smaller than the distance over which the wave-packets corresponding to different mass eigenstates separate because of the difference in their group velocities. This condition is satisfied if

\begin{equation}
    L \ll l_{\text{coh}} = \sigma_X\frac{v_{g_{ij}}}{\Delta v_{g_{ij}}}\;, \label{prop coherence}
\end{equation}
where $\sigma_X=\max \Big\{\sigma_{x}^S,\sigma_{x}^D\Big\}$ is the maximum wave-packet spatial width between the source (S) and the detector (D), and $\Delta v_{g_{ij}} = \abs{v_{g_i}-v_{g_j}}$.

The \emph{interaction coherence condition}, related to the coherence properties in the neutrino
production and detection processes \cite{Akhmedov:2009rb}, ensures that the two momentum distributions overlap in the domain of the integral. This condition reads
\begin{equation}\label{int coherence }
    \Delta E_{ij}\frac{\sigma_X}{v_{g_{ij}}} \ll 1\;.
\end{equation}

When conditions \eqref{prop coherence} and \eqref{int coherence } are satisfied, one can set $r_{ij}=1$ in the integral \eqref{oscillation probability standard wave-packet final} and neglect the term $\Delta E_{ij}$ inside the arguments of the distribution function. This leads to the standard neutrino oscillation formula typically derived within the plane wave assumption:
\begin{equation}\label{eq:oscillation standard}
     P(\beta\to\alpha) = \sum_{i,j} U^*_{\beta i} U_{\beta j} U^*_{\alpha j} U_{\alpha i}\,e^{i\phi_{ij}}\;,
\end{equation}
with $\phi_{ij}$ given by \eqref{phase factor}.

\section{Fundamental decoherence in neutrino oscillation}
\label{decoherence sec}

Fundamental decoherence causes a non-unitary evolution of the neutrino wave packet. This can be described by generalizing the Hamiltonian evolution via a Lindblad equation \cite{Lindblad1976},
\begin{equation}\label{Lindblad}
    \partial_t \rho = \mathfrak{L}\qty[\rho]\;,
\end{equation}
where the evolution operator is $\mathfrak{L}\qty[\rho] = -i [H,\rho] + \sum_k \qty(\mathcal{L}_k\rho \mathcal{L}_k^\dagger -\frac{1}{2}\pb{\mathcal{L}_k^\dagger \mathcal{L}_k}{\rho})$ and $\mathcal{L}_k$ are the Lindblad operators, that depend on the specific decoherence mechanism, as we discuss later in this section. Then $e^{-iHt}$ in \eqref{time evolution standard} is replaced by $e^{t\mathfrak{L}}$, yielding the density operator evolution: 

\begin{eqnarray}
    \rho(t) &=& e^{t \mathfrak{L}} \qty[\ketbra{\nu_\beta}{\nu_\beta}] = \sum_{i,j} U_{\beta i}^* U_{\beta j} \int \text{d}^3p \,\text{d}^3 q\, \psi_i\qty(\pmb{p})\psi_j^*\qty(\pmb{q}) e^{t\mathfrak{L}}\Big[\ketbra{\pmb{p}}{\pmb{q}}\otimes \ketbra{\nu_i}{\nu_j}\Big]\nonumber\\
    &=& \sum_{i,j} U_{\beta i}^* U_{\beta j} \int \text{d}^3p \,\text{d}^3 q\,\psi_i\qty(\pmb{p})\psi_j^*\qty(\pmb{q})e^{-i\qty[E_i\qty(\pmb{p})-E_j\qty(\pmb{q})]t}e^{-t\mathcal{L}_{ij}(\pmb{p},\pmb{q})} \ketbra{\pmb{p}}{\pmb{q}}\otimes \ketbra{\nu_i}{\nu_j}\;, \label{time evolution decohe}
\end{eqnarray}
where in the last line the Lindblad operators are taken to commute with the momentum operator, an assumption that is consistent with the specific decoherence models we consider in this work, and $\mathcal{L}_{ij}(\pmb{p},\pmb{q})$ is a model-dependent positive function of $\pmb{p}$ and $\pmb{q}$ resulting from the action of the Lindblad operators $\mathcal{L}_k$ on the $\ketbra{\pmb{p}}{\pmb{q}}$ eigenstates. Restricting to the one dimensional approximation, and following similar steps as in the previous section, we obtain an expression similar to \eqref{probbb}, with an additional damping exponential in the integral
\begin{eqnarray}
    &&P_{QG}(\beta\to\alpha;t) = \mathcal{N}\sum_{i,j} U^*_{\beta i} U_{\beta j} U^*_{\alpha j} U_{\alpha i} \,e^{i\Delta p_{ij}L} \int \text{d}p \,\text{d} q\,f_i^{(S)}(p) {f_j^{(S)}}^*(q) \nonumber \\ &&\cdot f_j^{(D)}(q+\delta_j) e^{i(p-q)L} {f_i^{(D)}}^*(p+\delta_i) e^{-i\qty[pv_{g_i}-qv_{g_j}+\Delta E_{ij}]t} e^{-t \mathcal{L}_{ij}(p+p_i,q+p_j)}\;,
\end{eqnarray}
where $P_{QG}(\beta\to\alpha;t)$ denotes the deformed probability. Averaging over time would produce a divergent probability, since $\mathcal{L}_{ij}(p,q)\geq 0\;\forall\,p,q$. This can be traced back to the fact that averaging over time in the presence of decoherence is inherently an ill-defined procedure, since decoherence processes break the time symmetry of quantum mechanical laws. The time interval $]-\infty,0[$ corresponds to an nonphysical "re-coherence", which is the cause of the divergent probability. We therefore directly enforce the equal-energy condition on wave packet components, as discussed in the previous section. Further substituting the remaining time variable in the damping factor with $\frac{L}{v_{g_{ij}}}$, the transition probability over a propagation distance $L$ reads
\begin{eqnarray}
    &&P_{QG}(\beta\to\alpha;L) = \mathcal{N}\sum_{i,j} U^*_{\beta i} U_{\beta j} U^*_{\alpha j} U_{\alpha i} \,e^{i \phi_{ij}} \frac{2\pi}{v_{g_j}}\inttt{p} f_i^{(S)}(p) {f_j^{(S)}}^*(r_{ij}\,p+\Delta E_{ij}/v_{g_j}) \cdot\nonumber \\ &&\cdot f_j^{(D)}(r_{ij}\,p+\Delta E_{ij}/v_{g_j}+\delta_j) {f_i^{(D)}}^*(p+\delta_i) e^{ip(1-r_{ij})L}e^{-D_{ij}\qty(p+p_i,\,r_{ij}p+p_j-v_{g_j}^{-1}\Delta E_{ij})}\;,\label{oscillation probability decohe wave-packet general}
\end{eqnarray}
where $D_{ij}(p,q)\ebd v_{g_{ij}}^{-1}L\,\mathcal{L}_{ij}(p,q)$ is the decoherence-induced damping factor. If the coherence conditions discussed at the end of the previous section are satisfied, the normalized transition probability reads 

\begin{equation}
    P_{QG}(\beta\to\alpha;L) = \sum_{i,j} U^*_{\beta i} U_{\beta j} U^*_{\alpha j} U_{\alpha i} \,e^{i \phi_{ij}} e^{-D_{ij}(p_i,p_j)}\;.\label{Oscillation probability decoh general}
\end{equation}
This general formula applies to any decoherence model described by Lindblad-type evolution, provided that the Lindblad operators commute with the momentum operator.
The effect of decoherence is to introduce a damping factor in the transition probability, such that when the decoherence effect is dominant no oscillation pattern in the neutrino propagation is observed.
In subsection \ref{decoherence models} we specialize this formula to different models of quantum-gravity-induced fundamental decoherence. 

\subsection{Two flavours analysis}
Most neutrino oscillation experiments are sensitive to the oscillation between two neutrino flavours, depending on their sensitivity to different mass squared differences. In the two-flavour case, the matrix $U$ in \eqref{Oscillation probability decoh general} is parameterized in terms of the mixing angle $\theta$ as
\begin{equation}
    U = \begin{pmatrix}
\cos \theta & \sin \theta \\
-\sin \theta & \cos \theta
\end{pmatrix}\;,
\end{equation}
and the oscillation phase \eqref{phase factor} reduces to $\phi= -L\frac{\Delta m^2}{2\,p}$, with $\Delta m^2$ the mass squared difference and $p$ the average momentum.
Then the standard transition probability \eqref{eq:oscillation standard} takes the well-known form
\begin{equation}\label{eq:standard osc 2 flavours}
    P_{\text{std}}(\alpha\to\beta) = \sin^2{2\theta}\sin^2{\frac{\phi}{2}}
\end{equation}
and the transition probability accounting for fundamental decoherence, \eqref{Oscillation probability decoh general}, reduces to

\begin{equation}\label{Probability decoherence two flavour}
    P_{QG}(\alpha\to\beta) =  \sin^2{2\theta}\qty[\frac{1}{2}\qty(1-e^{-D})+e^{-D}\sin^2{\frac{\phi}{2}}] = e^{-D} P_{\text{std}}(\alpha\to\beta)+\frac{1}{2}\qty(1-e^{-D})\sin^2{2\theta}\;,
\end{equation}
with $D= v_g^{-1}L\mathcal{L}(p_i,p_j)$, where $v_g = \frac{p}{E} = \frac{p}{\sqrt{p^2+m^2}}$ is the average group velocity of the two mass eigenstates, $m$ is the average mass, $p_i,p_j$ are the mean momenta of such states and $p$ is the average momentum.

For the analysis reported in the following sections, it is useful to consider the survival probability 
\begin{equation}\label{Survival probability two flavours}
    P_{QG}(\alpha\to\alpha) \equiv 1- P_{QG}(\alpha\to\beta)= e^{-D} P_{\text{std}}(\alpha\to\alpha) + \qty(1-e^{-D})\qty(1-\frac{1}{2}\sin^2 2\theta)\;,
\end{equation}
where
\begin{equation}\label{standard survival}
    P_{\text{std}}(\alpha\to\alpha) = 1 - P_{\text{std}}(\alpha\to\beta) = 1-\sin^2{2\theta}\sin^2{\frac{\phi}{2}}\;.
\end{equation}
This is consistent with the result reported in \cite{Lisi:2000zt}, where however different possible forms for the damping factor were considered with respect to this work.

\subsection{Damping factor for different decoherence models}\label{decoherence models}

In this section we specialize the general formulas \eqref{oscillation probability decohe wave-packet general} and \eqref{Survival probability two flavours}, that account for the effects of a Lindlbad-type evolution on neutrino oscillations, 
to a number of decoherence mechanisms that find support in quantum gravity research. Specific decoherence models can be distinguished according to the functional dependence of the damping factor on the neutrino energies and masses.
In the following, we focus on four models, whose properties are summarized in \cref{table models}, in which fundamental decoherence is induced by different possible features of quantum spacetime: non-commutative spacetime \cite{Arzano:2022nlo}, stochastic fluctuations of a minimal length \cite{Petruzziello:2020wkd}, stochastic fluctuations of the metric around Minkowski spacetime \cite{Breuer:2008rh}, and the interaction with a thermal background of gravitons \cite{Blencowe:2012mp}. 

Starting from the Lindblad operators characterizing the evolution of quantum systems in each of these models, the damping factor $D_{ij}(p,q)$ appearing in the oscillation probability \eqref{oscillation probability decohe wave-packet general} and the resulting oscillation probability \eqref{Oscillation probability decoh general} can be computed as described in the previous section. Notice that, while in the original papers defining the various models decoherence is assumed to be suppressed by the Planck scale $E_P$, in the following we replace this with a generic quantum gravity scale $E_{QG}$, to be constrained by observations. 

Decoherence induced by non-commutative spacetime \cite{Arzano:2022nlo} leads to the transition probability 

\begin{equation}
    P(\beta\to\alpha)  = \sum_{i,j} U^*_{\beta i} U_{\beta j} U^*_{\alpha j} U_{\alpha i} \,e^{i \phi_{ij}} e^{-\frac{L}{8v_{g_{ij}}E_{QG}\,p^2_{ij}}\qty(\Delta m_{ij}^2)^2}\;.\label{Oscillation probability decoh arzano}
\end{equation}
In the two-flavour approximation, the damping factor reads
\begin{equation}\label{damping factor 2 flavours arzano}
    D = \frac{L}{8v_gE_{QG} \,p^2}(\Delta m^2)^2\;.
\end{equation}

Decoherence induced by stochastic fluctuations of the metric \cite{Breuer:2008rh} results in the transition probability 
    \begin{equation}
        P(\beta\to\alpha) = \sum_{i,j} U^*_{\beta i} U_{\beta j} U^*_{\alpha j} U_{\alpha i} \,e^{i \phi_{ij}} e^{-\frac{L\,E_{ij}^6 \qty(\Delta m^2_{ij})^2}{4v_{g_ij}E_{QG}\,m_i^4m_j^4}}\;.\label{Oscillation probability decoh Breuer}
    \end{equation}
Then the two-flavour damping factor is \begin{equation}\label{damping factor 2 flavours analysis Breuer}
        D = \frac{L\,E^6 \qty(\Delta m^2)^2}{4v_{g}E_{QG}\,m_i^4m_j^4}\;.
    \end{equation}

Depending on whether we consider the quasi-degenerate mass eigenstates $m_1$ and $m_2$ or the mass eigenstates $m_1$ (or $m_2$) and $m_3$, we can write the product of the masses in terms of the average mass or the mass squared difference. If the two mass eigenstate of interest are $m_1$ and $m_2$, as it happens to be the case for reactor neutrinos, we have $m_im_j \sim m^2$, $m$ being the average mass. If the mass eigenstates of interest are $m_1$ (or $m_2$) and $m_3$, as it happens to be the case for atmospheric neutrinos, then  $m_im_j = m_{\text{min}}\sqrt{m^2_{\text{min}}+\Delta m^2}$, $m_{\text{min}}$ being the minimum mass between the two mass eigenstates.
    
The decoherence process induced by a fluctuating minimal length \cite{Petruzziello:2020wkd} gives the transition probability
    \begin{equation}
         P(\beta\to\alpha) = \sum_{i,j} U^*_{\beta i} U_{\beta j} U^*_{\alpha j} U_{\alpha i} \,e^{i \phi_{ij}} e^{-\frac{16L}{v_{g_{ij}}E_{QG}^5}E_{ij}^4\qty(\Delta m_{ij})^2}\;,\label{Oscillation probability decoh Petruziello}
    \end{equation}
and the damping factor in the two-flavour approximation 
    \begin{equation}\label{damping factor 2 flavours Petruzziello}
        D=\frac{16L}{v_{g}E_{QG}^5}E^4\qty(\Delta m)^2\;.
    \end{equation}
Depending on whether we consider the quasi-degenerate mass eigenstates or not, we can approximate the mass difference squared with the squared mass difference or write the former in terms of the latter. For the quasi-degenerate mass eigenstates, we have $\qty(\Delta m)^2 \sim \Delta m^2$. For the non degenerate mass eigenstates we have instead $\qty(\Delta m)^2 = \qty(-m_{\text{min}}+\sqrt{m^2_{\text{min}}+\Delta m^2})^2$.
    
A thermal background of gravitons \cite{Blencowe:2012mp} yields the exponential factor 
\begin{equation}
    \exp{-t\frac{k_BT}{2\,E_{QG}^2}\qty(E_i(p)-E_j(q))^2}
\end{equation}
in the evolved density operator \eqref{time evolution decohe}. Therefore, this model predicts no contribution to the probability after the equal-energy condition is imposed and  has no observable effect on neutrino oscillations. Therefore,  neutrino oscillations cannot provide a  phenomenological test ground for this model. We summarize the results described above in \cref{table models}. We note that, among the models we consider, the only one that can be recast in terms of the phenomenological models analysed in \cite{DeRomeri:2023dht} is the first one in \cref{table models}. Specifically, the third model in \cref{table models}, that turns out to be the most interesting one for our purposes, is not included in the analysis of \cite{DeRomeri:2023dht}.

\begin{table}
    \centering
{\tabulinesep=1.2mm
        \begin{tabu}{|l|l|l|l|l}
        \cline{1-4}
       \small\bf{Model}&{\small\bf{Physical source of decoherence}} &{\small\bf{Lindblad operators}}& {\small\bf{Damping factors}}& \\ \cline{1-4}A)&
       {\footnotesize Deformation of symmetries \cite{Arzano:2022nlo}}  & $\mathcal{L}_j = \frac{1}{\sqrt{E_{QG}}}P_j$ &  $D = \frac{L\qty(\Delta m^2)^2}{8v_{g}E_{QG}\,p^2}$&\\ \cline{1-4}
       B)&{\footnotesize  Metric perturbations  \cite{Breuer:2008rh} } & $\mathcal{L} = \frac{\pmb{P}^2}{2m\sqrt{E_{QG}}}$ & $D = \frac{L\,E^6 \qty(\Delta m^2)^2}{4v_{g}E_{QG}\,m_i^4m_j^4}$&\\ \cline{1-4}
        C)&{\footnotesize Fluctuating minimal length \cite{Petruzziello:2020wkd}} & $\mathcal{L} = \frac{4\sqrt{2}m}{\sqrt{E_{QG}^5}}H^2$ & $D = \frac{16LE^4\qty(\Delta m)^2}{v_{g}E_{QG}^5}$ &\\ \cline{1-4}
        D)&{\footnotesize Gravitons at temperature $T$ \cite{Blencowe:2012mp}} &$\mathcal{L} = \frac{\sqrt{k_B T}}{E_{QG}}H$ &  $D=0$&\\  \cline{1-4}
        \end{tabu}}
    \caption{Lindblad operators and damping factors for different quantum gravity-induced decoherence models. $P_j$ is the $j-$th component of the momentum operator, $E_{QG}$ is the quantum gravity scale, $k_B$ is the Boltzmann constant, $m$ is the mass and $H$ is the free Hamiltonian. The Lindlbad operator for the model \cite{Breuer:2008rh} is computed within a relativistic version of the model, see \cref{appendix relativistic model}.} 
    \label{table models}
\end{table}

\section{Experimental constraints on fundamental decoherence}\label{sec:experiments}

Different neutrino oscillation experiments, involving astrophysical, atmospheric, solar or reactor neutrino, span a wide range of propagation lengths and energy. 
In this section, we start by discussing the potential sensitivity to decoherence of these classes of neutrino oscillation experiments. This depends on the specific model of fundamental decoherence that is considered, since the damping factors that are derived from each model have different dependencies on the neutrino energies and masses.
We show that only the decoherence induced by metric perturbations \cite{Breuer:2008rh} may affect neutrino oscillations to an observable level, assuming that the decoherence effect is suppressed by the Planck energy scale, and we use atmospheric and reactor experiments to establish constraints on the model.

\subsection{Sensitivity of different neutrino oscillation experiments to fundamental decoherence}

Before considering the sensitivity of neutrino oscillation experiments to specific fundamental decoherence models, two general considerations are in order.

The standard neutrino oscillations pattern, governed by the phase factor $\phi$ of \eqref{phase factor}, depends on the ratio $\frac{E}{L}$: when $\frac{E}{L}\sim \Delta m^2$ oscillations are visible, while $\frac{E}{L}\ll \Delta m^2$ is the fast oscillation regime, where it is only possible to measure the averaged probability. In this fast oscillation regime, the standard survival probability for a two-flavour oscillation \eqref{eq:standard osc 2 flavours} reads  
\begin{equation}
    \expval{P_{\text{std}}(\alpha\to\alpha)} = 1-\frac{1}{2}\sin^2 2\theta\;.
\end{equation}
When accounting for decoherence, the average over the oscillations in \eqref{Survival probability two flavours} gives the same result as in the standard case
\begin{equation}\label{Average survival probability}
    \expval{ P_{QG}(\alpha\to\alpha)} = \expval{P_{\text{std}}(\alpha\to\alpha)} = 1-\frac{1}{2}\sin^2 2\theta\;,
\end{equation}
thus Lindblad-type decoherence has no observable effects in this regime.
This is the regime characterizing solar neutrino oscillations, that are therefore not relevant for this kind of decoherence studies.

Because the damping factor in the oscillation probability, $D_{ij}$, depends linearly on the propagation length $L$ (see \cref{table models}), one would naively expect that the best setup for testing fundamental decoherence is provided by astrophysical neutrino observations, for which $L$ is the largest.
However, in this regime the fundamental decoherence mechanism competes with the standard decoherence that is induced by the the difference in propagation velocities of different neutrino mass eigenstates, discussed in \cref{standard oscillations section}. For illustrative purposes, we discuss this by focusing on the model in \cite{Arzano:2022nlo}, for which the damping factor is $D_{ij}(p,q)= \frac{L}{2v_{g_{ij}}E_{QG}}\qty(p-q)^2$, but similar arguments hold for any Lindblad-type decoherence that leads to the oscillation probability \eqref{oscillation probability decohe wave-packet general}. We also consider a simple case in which the shape factors are Gaussian distributions with the same variance at the source and the detector for all the mass eigenstates, with $\delta_i=\delta_j=0$ and $\Delta E_{ij} = 0$. Specifically, we set
\begin{equation}\label{shape factors}
    f_i^{(S)}(p) \propto e^{-\frac{p^2}{2\sigma_p^2}}\;,
\end{equation}
where the normalization is unimportant since it is absorbed in the factor $\mathcal{N}$ in \eqref{oscillation probability decohe wave-packet general}. With these choices, and approximating $v_{g_j}$ with $v_{g_{ij}}$, eq. \eqref{oscillation probability decohe wave-packet general} yields
\begin{eqnarray}
    P(\beta\to\alpha) &=& \mathcal{N} 2\pi\sqrt{2\pi}\sum_{i,j} \Bigg\{U^*_{\beta i} U_{\beta j} U^*_{\alpha j} U_{\alpha i} \, e^{i \phi_{ij}\qty[1-\frac{L(1-r_{ij})^2\sigma_p^2}{2E_{QG}(1+r_{ij}^2)v_{g_{ij}}+L(1-r_{ij})^2\sigma_p^2}]}\nonumber \\ &\cdot&\frac{e^{-\frac{L^2(1-r_{ij})^2v_{g_{ij}}\sigma_p^2+\frac{2(1+r_{ij}^2)\phi_{ij}^2}{LE_{QG}}}{4(1+r_{ij}^2)v_{g_{ij}}+\frac{2L(1-r_{ij})^2\sigma_p^2}{E_{QG}}}}}{\sqrt{v_{g_{ij}}\qty(2(1+r_{ij}^2)v_{g_{ij}}+\frac{L(1-r_{ij})^2\sigma_p^2}{E_{QG}})}}\Bigg\}\;.\label{Oscillation probability decohe symm gaussian}
\end{eqnarray}
The exponential damping factor in \eqref{Oscillation probability decohe symm gaussian} takes two contributions: one, which vanishes when $E_{QG}\to \infty$, is due to the Lindblad deformation of the evolution equation; the other one, proportional to $(1-r_{ij})^2$, is due to the velocity difference between mass eigenstates. Even though the velocity difference is rather small, and its effect can be partially controlled by decreasing $\sigma_p$ through suitable detection techniques \cite{Giunti:1997wq}, the decoherence it induces is strong enough to wash out oscillations over propagation length scales typical of astrophysical neutrinos. Therefore, the effects of quantum gravity-induced decoherence are not observable.

For atmospheric and reactor neutrinos the standard decoherence effect we just discussed is negligible. Moreover, the full oscillation pattern is visible and not averaged out, as is instead the case for solar neutrinos. Therefore, in the following we focus on this kind of neutrino observations to study fundamental decoherence. 

We start by computing, for each of the models in \cref{table models},  the theoretical ranges for neutrino energies that would produce significant decoherence effects in oscillations when using the mass and length parameters typical of reactor and atmospheric neutrino experiments, and assuming $E_{QG}=E_P\sim 10^{19}\;\text{GeV}$. For simplicity, we work in a two-flavour setup and we require that the corresponding damping factor satisfies the condition $D\gtrsim 1$, which ensures that the decoherence effect is large enough to affect the oscillation pattern \cite{Christian:2005qa}. We then compare these energy ranges with the realistic energies characterizing reactor and atmospheric neutrinos, in order to identify the models that can be effectively constrained with this kind of observations.


\begin{table}
    \centering
{\tabulinesep=1.2mm
        \begin{tabu}{|l|l|l|}
        \cline{1-3}
       {\small\bf{Model}} & {\footnotesize \bf{Observability window for $m=10^{-2}$ eV}} &{\footnotesize \bf{Observability window for $m=1$ eV}} \\ \cline{1-3}
       {\small  A)} & $p\lesssim 10^{-8}\,\text{eV}$ & $p\lesssim 10^{-8}\; \text{eV}$\\ \cline{1-3}
           {\small  B)} &  $p\lesssim 10^{-22}\;\text{eV}$ or $p \gtrsim 100 \; \text{eV}$ & $p\lesssim 10^{-23}\;\text{eV}$ or $p \gtrsim 10^4 \; \text{eV}$\\ \cline{1-3}
        {\small  C)} & $p\lesssim 10^{-147}\;\text{GeV}$ or $p\gtrsim 10^{24}\,\text{GeV}$ & $p\lesssim 10^{-143}\;\text{GeV}$ or $p\gtrsim 10^{24}\,\text{GeV}$\\ \cline{1-3}
        {\small  D)} &  // & // \\  \cline{1-3}
        \end{tabu}}
    \caption{Observability windows for the decoherence models listed in \cref{table models} using reactor neutrino experiments, for different values of the neutrino mass. Details of the computations are provided in the main text. Note that when two ranges are shown for $p$, this comes from solving the quadratic equation that comes from the condition $D \gtrsim 1$.}
    \label{table results}
\end{table}

For long-baseline reactor experiments, we take as reference propagation distance $L\sim 10^5\;\text{m}$ 
and we set $\Delta m^2 \sim \qty(\Delta m)^2 \sim 10^{-5} \;\text{eV}^2$ \cite{KamLAND:2008dgz}, since such experiments are sensitive to the mass squared difference between the quasi-degenerate mass eigenstates $m_1$ and $m_2$. The damping factors of the various decoherence models also depend on the neutrino mass value, for which we take two possible reference values, $m\sim 10^{-2}\;\text{eV}$ and $m \sim 1\;\text{eV}$, the maximum value allowed by cosmological data \cite{DiValentino:2021imh}. Once these parameters are fixed, the requirement $D\gtrsim 1$ identifies a range in neutrino momentum $p$ for each decoherence model. These are listed in \cref{table results}. Considering that the typical energy of reactor neutrinos, such as those detected by the KamLAND experiment \cite{KamLAND:2008dgz}, is in the range $[1,10]\; \text{MeV}$ \cite{KamLAND:2008dgz}, only decoherence induced by stochastic metric fluctuations \cite{Breuer:2008rh} might be constrained significantly using this kind of data. 

A similar analysis can be applied to atmospheric neutrino oscillation experiments, that are sensitive to the mass squared difference between eigenstates $m_{1,2}$ and $m_3$, so that $\Delta m^2 \sim 10^{-3}\;\text{eV}^2$ \cite{Super-Kamiokande:2004orf}. We assess the sensitivity of these experiments considering again two possible values of the lowest-mass eigenstate, $m_{\text{min}} \sim 10^{-2}\;\text{eV}$ and $m_{\text{min}}\sim 1\;\text{eV}$. Atmospheric neutrinos can propagate over a wide range of distances, spanning several orders of magnitude, from $10\;\text{Km}$ to $10^4\;\text{Km}$ \cite{Super-Kamiokande:2004orf}. For the scope of estimating the sensitivity of this kind of observations we take the minimum and the maximum possible values. The observability windows are reported in \cref{windows_atmo_table} and \cref{windows_atmo_table_2} respectively. Again, we see that only decoherence induced by stochastic metric fluctuations \cite{Breuer:2008rh} might be constrained significantly using atmospheric neutrino oscillation data. 

Summarizing the results of this subsection, solar and astrophysical neutrinos cannot be used to test Lindblad-type decoherence. Reactor and atmospheric neutrino experiments can in principle be used to this aim. However, not all models predict a level of decoherence that might be detected by these experiments, if the energy scale governing the strength of decoherence is set to the Planck scale. Among the four models we considered, only the one where decoherence is induced by metric fluctuations can be meaningfully constrained using these experiments. In the following subsections we derive the constraints on the energy scale associated to this model.

Before we conclude this section, we would like to comment on the concerns expressed in \cite{Adler:2000vfa}, where the possibility to test quantum gravity-induced decoherence with neutrino oscillation experiments was questioned. The quantum gravity model considered in \cite{Adler:2000vfa} (see also \cite{Lisi:2000zt}) had a similar form as the one of \cite{Petruzziello:2020wkd} (since the Lindblad operator depends on the energy). And indeed we find that the damping factor associated to this model is heavily suppressed, in accordance with the estimates provided in \cite{Adler:2000vfa}. For other models, such as the one in \cite{Breuer:2008rh}, similar arguments as those used in \cite{Adler:2000vfa} do not lead to such a large suppression of the damping factor, and in fact lead to estimates of the damping factor consistent with those we report in \cref{table models}.

\begin{table}
    \centering
{\tabulinesep=1.2mm
        \begin{tabu}{|l|l|l|}
        \cline{1-3}
       {\small\bf{Model}} & {\footnotesize \bf{Observability window for $m_{\text{min}}=10^{-2}$ eV}} &{\footnotesize \bf{Observability window for $m_{\text{min}}=1$ eV}} \\ \cline{1-3}
       {\small  A)} & $p\lesssim 10^{-8}\,\text{eV}$ & $p\lesssim 10^{-7}\; \text{eV}$\\ \cline{1-3}
           {\small  B)} &  $p\lesssim 10^{-19}\,\text{eV}$ or $p\gtrsim 10^{2}\,\text{eV}$  & $p\lesssim 10^{-20}\,\text{eV}$ or $p\gtrsim 10^{3}\,\text{eV}$\\ \cline{1-3}
        {\small  C)} & $p\lesssim 10^{-145}\;\text{GeV}$ or $p\gtrsim 10^{23}\,\text{GeV}$ & $p\lesssim 10^{-141}\;\text{GeV}$ or $p\gtrsim 10^{24}\,\text{GeV}$\\ \cline{1-3}
        {\small  D)} &  // & // \\  \cline{1-3}
        \end{tabu}}
    \caption{Observability windows for the decoherence models listed in \cref{table models} using atmospheric neutrino experiments, for different values of the neutrino mass $L=10\,\text{Km}$. Details of the computations are provided in the main text.}
    \label{windows_atmo_table}
\end{table}

\begin{table}
    \centering
{\tabulinesep=1.2mm
        \begin{tabu}{|l|l|l|}
        \cline{1-3}
       {\small\bf{Model}} & {\footnotesize \bf{Observability window for $m_{\text{min}}=10^{-2}$ eV}} &{\footnotesize \bf{Observability window for $m_{\text{min}}=1$ eV}} \\ \cline{1-3}
       {\small  A)} & $p\lesssim 10^{-7}\,\text{eV}$ & $p\lesssim 10^{-6}\; \text{eV}$\\ \cline{1-3}
           {\small  B)} &  $p\lesssim 10^{-16}\,\text{eV}$ or $p\gtrsim 10\,\text{eV}$  & $p\lesssim 10^{-18}\,\text{eV}$ or $p\gtrsim 10^{3}\,\text{eV}$\\ \cline{1-3}
        {\small  C)} & $p\lesssim 10^{-142}\;\text{GeV}$ or $p\gtrsim 10^{22}\,\text{GeV}$ & $p\lesssim 10^{-138}\;\text{GeV}$ or $p\gtrsim 10^{23}\,\text{GeV}$\\ \cline{1-3}
        {\small  D)} &  // & // \\  \cline{1-3}
        \end{tabu}}
    \caption{Observability windows for the decoherence models listed in \cref{table models} using atmospheric neutrino experiments, for different values of the neutrino mass and $L=10^4\,\text{Km}$. Details of the computations are provided in the main text.}
    \label{windows_atmo_table_2}
\end{table}

\subsection{Constraints from long-baseline reactor neutrinos}\label{analysis KAM}

\begin{figure}
    \centering
    \includegraphics[width=0.9\textwidth]{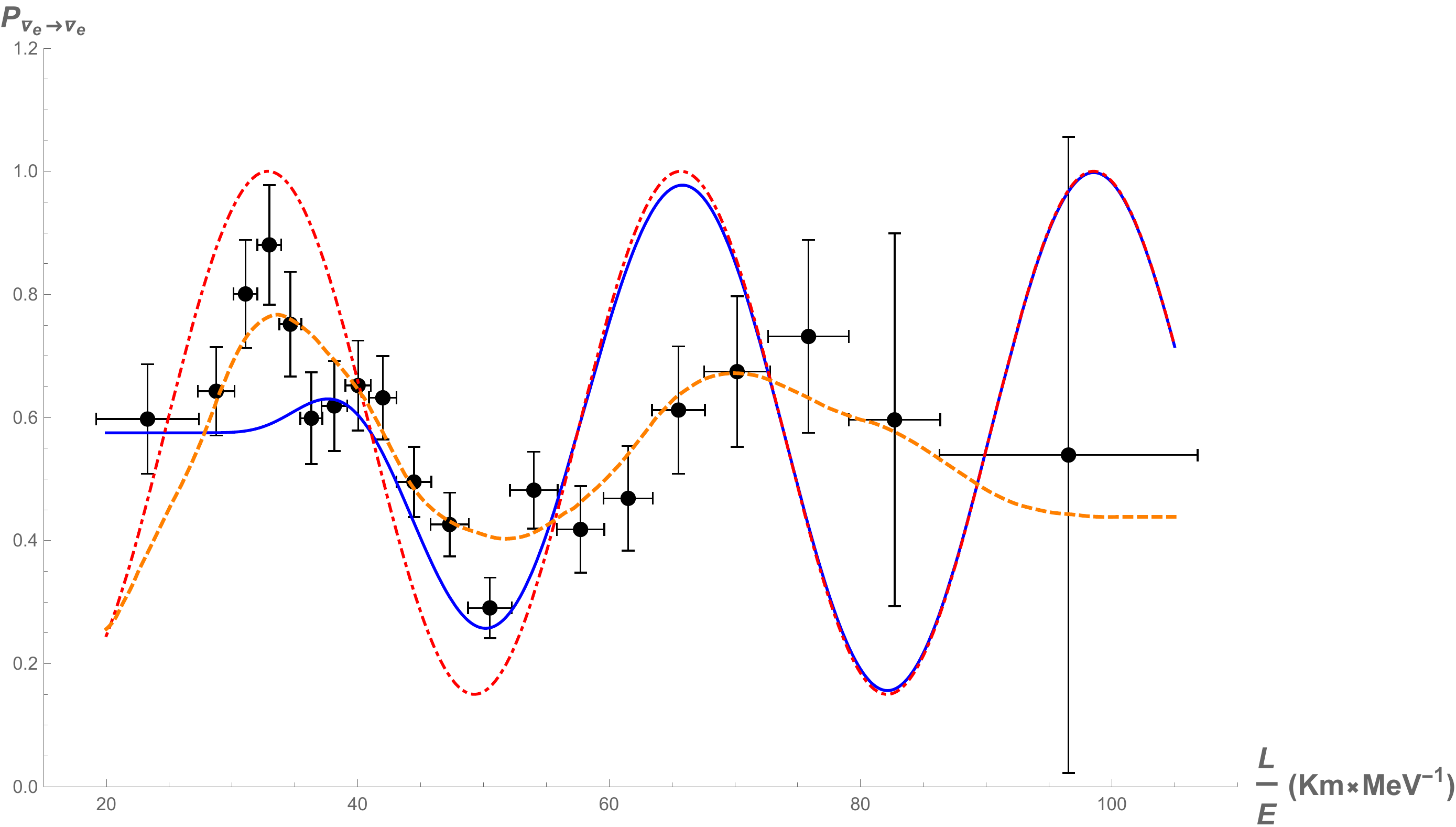}
    \caption{We plot the standard survival probability (red dot-dashed curve) in vacuo, given by \eqref{standard survival}, the quantum gravity-deformed probability with damping factor \eqref{damping factor 2 flavours analysis Breuer} (blue solid line), the data points from KamLAND experiment \cite{KamLAND:2008dgz} (black dots with error bars) and the KamLAND best fit for such data (orange dashed curve). The oscillation parameters are set to the PDG values in \cite{Workman:2022ynf} $\sin^2 2\theta = 0.85$ and $\Delta m^2 = 7.53\cdot 10^{-5}\text{eV}^2$. The damping factor \eqref{damping factor 2 flavours analysis Breuer} is computed by setting $L=180\;\text{Km}$, $E_{QG}=10^{34}\;\text{GeV}$ and $m=1\;\text{eV}$. }
    \label{Plot KamLAND}
    \end{figure}

\begin{figure}
    \centering
    \includegraphics[width=0.9\textwidth]{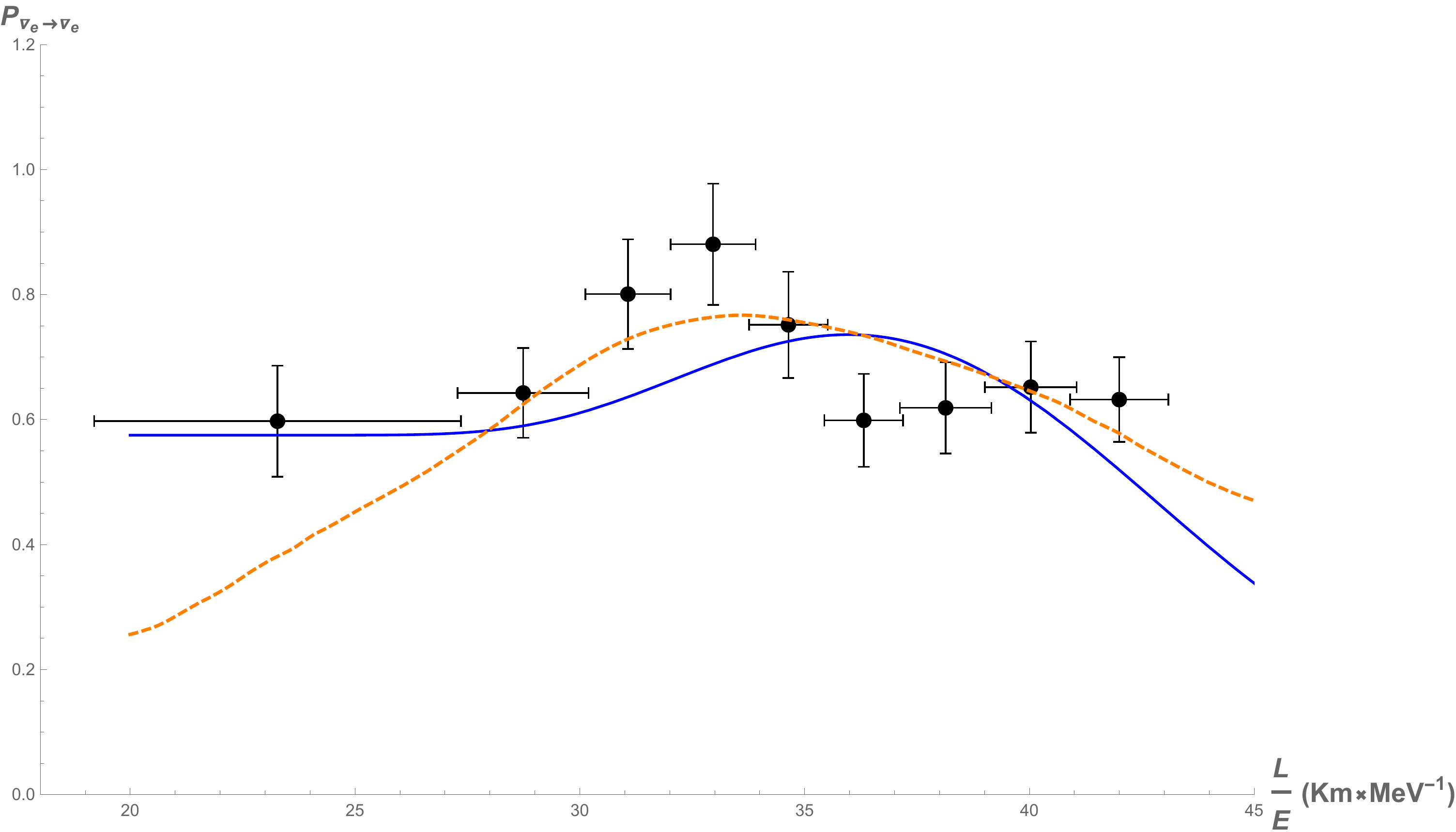}
    \caption{We plot the quantum gravity-deformed probability with damping factor \eqref{damping factor 2 flavours analysis Breuer} (blue solid line) with the value of $E_{QG}=2.6 \cdot 10^{34}\;\text{GeV}$ corresponding to the upper limit we set as described in the main text, the data points from KamLAND experiment \cite{KamLAND:2008dgz} in the range $\frac{L}{E} \in [10,45]\;\text{Km}\; \text{MeV}^{-1}$ (black dots with error bars) and the KamLAND best fit for the full data set reported in \cref{Plot KamLAND} (orange dashed curve). The oscillation parameters and the other decoherence parameter are the same as for \cref{Plot KamLAND}.} 
    \label{Plot KamLAND reduced} 
\end{figure}

We base our analysis on the data used by the KamLAND collaboration for precision measurements of neutrino oscillation parameters concerning the mass eigenstates $m_1$ and $m_2$ \cite{KamLAND:2008dgz}.
Within this two-flavour approximation, the survival probability accounting for Lindblad-type decoherence is \eqref{Survival probability two flavours}, and the damping factor $D$ for decoherence induced by metric fluctuations \cite{Breuer:2008rh} is given by \eqref{damping factor 2 flavours analysis Breuer}\footnote{Given the energies at play in this regime, the velocity factor is $v_g\simeq1$.}.
As can be seen in \cref{Plot KamLAND}, neutrino oscillations in the KamLAND experiment are damped with respect to the pattern of a neutrino beam traveling in vacuo and with a fixed baseline. This is expected, since several reactors are involved in the experiment, placed at different distances from the detectors, and the neutrino beam travels through matter. Including these effects would require a more refined analysis and a full knowledge of the KamLAND experiment. However, such a detailed modelling goes beyond the scopes of this work, since it is not needed when the goal is to provide a conservative bound on quantum gravity-induced decoherence. To this aim, in fact, it is sufficient to ask that decoherence is not as strong as to dampen oscillations on its own more than what the known damping effects do. Indeed, if this was not the case, the damping stemming from quantum-spacetime effects, combined with this standard damping, would completely quench the oscillations.
In \cref{Plot KamLAND} we also plot the survival probability \eqref{Survival probability two flavours}, computed taking into account the appropriate values for experimental parameters as detailed in the figure caption. The only free parameter is the quantum gravity scale $E_{QG}$, and to produce the plot we fixed it to a reference value $E_{QG} = 10^{24}\;\text{GeV}$ for illustrative purposes. We can see that decoherence effects are stronger at higher energies (lower values of $\frac{L}{E}$) while at lower energies the deformed survival probability approaches the standard one in vacuo. Therefore, the most sensitive regime to test decoherence is the one corresponding to the first oscillation period, so we use data points in the range $\frac{L}{E}\in [10,45]\;\text{Km}\; \text{MeV}^{-1}$ to provide a constraint on the model, see \cref{Plot KamLAND reduced}. This restriction is justified by the fact that if the quantum spacetime-induced damping had observable effects in the remaining part of the $\frac{L}{E}$ interval, it would also have a much stronger effect on the oscillation pattern in the considered $\frac{L}{E}$ interval, thus being incompatible with the data in that regime.

Our constraint is derived by asking that the difference between the $\chi^2$ obtained by fitting the deformed probability \eqref{Survival probability two flavours}, with D given by \eqref{damping factor 2 flavours analysis Breuer}, to KamLAND data and the $\chi^2$ corresponding to KamLAND best fit line is not larger than $2.7$, corresponding to a $90\%$ confidence level for a difference of $1$ in the counting of degrees of freedom for the two fits. In the expression for the damping factor \eqref{damping factor 2 flavours analysis Breuer}, we set the oscillation parameters to be the values reported by the Particle Data Group (PDG) in \cite{Workman:2022ynf}. We also set $L = 180\;\text{Km}$ as done in the original oscillation analysis by the KamLAND collaboration \cite{KamLAND:2008dgz}. Moreover, since the constraint on the quantum gravity scale is stronger for smaller neutrino masses (see \eqref{damping factor 2 flavours analysis Breuer}), a conservative bound is found by setting $m=1\;\text{eV}$, namely:
\begin{equation}
    E_{QG}\geq 2.6\cdot 10^{34}\;\text{GeV}\;.
    \end{equation}

\subsection{Constraints from atmospheric neutrinos}\label{analysis SK}

\begin{figure}
    \centering
    \includegraphics[width=0.9\textwidth]{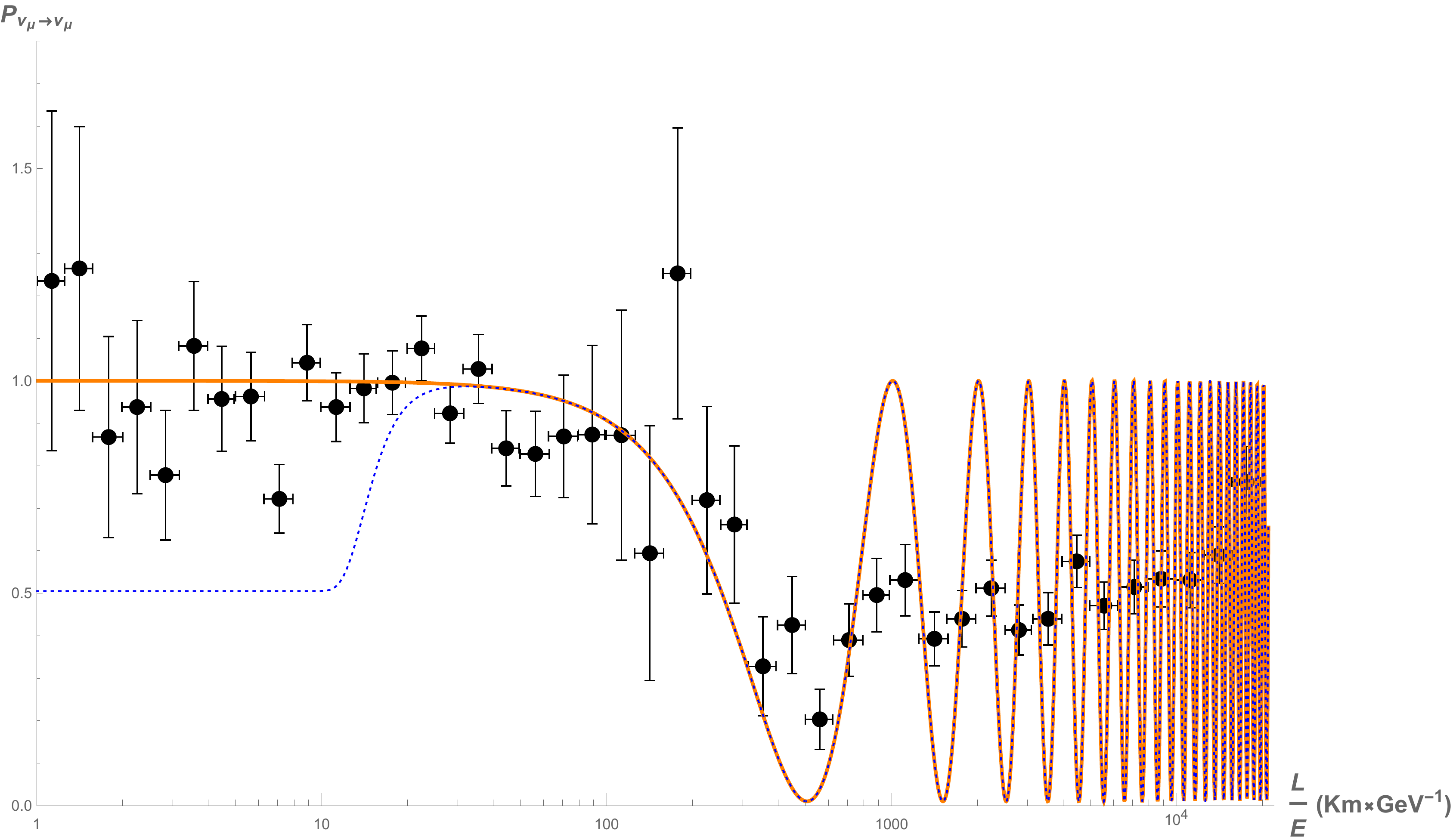}
    \caption{We plot the standard oscillation probability (solid orange curve), the quantum gravity-deformed probability (dotted blue curve) with damping factor \eqref{damping factor 2 flavours analysis Breuer}, with $m_{\text{min}}= 1\,\text{eV}$, $L= 10\;\text{Km}$ and $E_{QG} = 10^{49}\,\text{GeV}$, chosen for illustrative purposes, and the data collected in \cite{Super-Kamiokande:2004orf}. The values of the oscillation parameters used for the plot are those reported in the PDG \cite{Workman:2022ynf}, $\Delta m_{23}^2 = 2.45 \cdot 10^{-3}\;\text{eV}^2$ and $\sin^2 2\theta_{23} = 0.99$. The $\frac{L}{E}$ axis is in logarithmic scale.}
    \label{Plot SK}
\end{figure}

We use the data collected by the Super-Kamiokande collaboration and reported in \cite{Super-Kamiokande:2004orf}. The relevant $\frac{L}{E}$ range is $[1,10^4]\;\text{Km}\;\text{GeV}^{-1}$, so that the observed oscillation pattern is sensitive to the parameters relative to the mass eigenstates $m_1$ and $m_3$. As can be seen in \cref{Plot SK}, the fast oscillations regime starts for $\frac L E \gtrsim 10^3 \;\text{Km}\; \text{GeV}^{-1}$. In this regime, the data only probe the average oscillations and the averaged quantum gravity-deformed survival probability converges to the standard one, see \eqref{Average survival probability}. For this reason, we limit our analysis to data in the range $\frac{L}{E}\in [1,10^3]\;\text{Km}\;\text{GeV}^{-1}$, where the oscillation pattern is visible and decoherence produces observable effects.
We compare the $\chi^2$ obtained by fitting the data with the deformed probability \eqref{Survival probability two flavours}, with $D$ given by \eqref{damping factor 2 flavours analysis Breuer}\footnote{Given the energies at play in this regime, the velocity factor is $v_g\simeq1$.}, and the $\chi^2$ corresponding to the standard oscillation pattern with oscillation parameters taken from the PDG \cite{Workman:2022ynf}. We require that the difference between the two $\chi^2$ is less than $2.7$, corresponding to a $90\%$ confidence level for the $1$ degree of freedom difference between the two models. A conservative constraint is found by setting $m_{\text{min}}=1\;\text{eV}$ in \eqref{damping factor 2 flavours analysis Breuer}, since the constraint on the quantum gravity scale would be stronger with smaller values of the mass. Moreover, each data point in \cref{Plot SK} results from oscillation over a distance that spans several orders of magnitude, roughly from $L=10\;\text{Km}$ to $L=10^4$\;\text{Km}. 
Because the constraint on the quantum gravity scale is stronger for higher travel distances (see \eqref{damping factor 2 flavours analysis Breuer}), we take a conservative approach again and set $L=10\;\text{Km}$ in our analysis. With this, we find

\begin{equation}\label{constraint Kamio}
    E_{QG}\geq 2.5\cdot 10^{55}\;\text{GeV}\;.
\end{equation}

\begin{figure}
    \centering
    \includegraphics[width=0.89\textwidth]{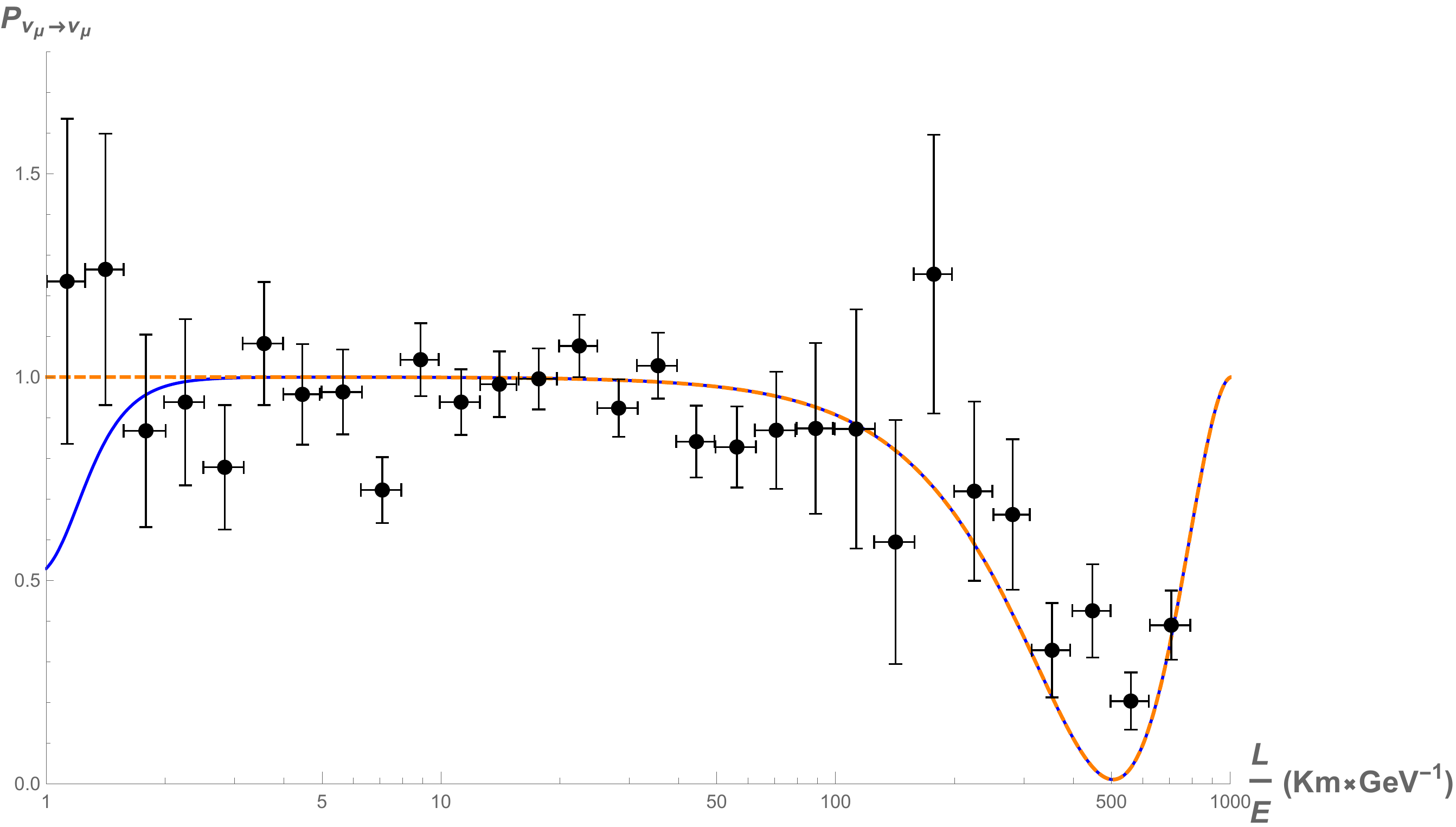}
    \caption{We plot the standard oscillation probability (solid orange curve), the quantum gravity-deformed probability (blue solid curve) with damping factor \eqref{damping factor 2 flavours analysis Breuer}, with $E_{QG} = 2.5\cdot 10^{55}\,\text{GeV}$, corresponding to our upper limit \eqref{constraint Kamio}, and the data collected in \cite{Super-Kamiokande:2004orf} in the range $\frac{L}{E} \in [1,10^3]\;\text{Km}\; \text{GeV}^{-1}$. The values of the oscillation parameters and the other decoherence parameters are the same as in \cref{Plot KamLAND}. }
    \label{Plot SK reduced}
\end{figure}

\section{Conclusions}

We showed that fundamental decoherence, originating from quantum properties of spacetime, can alter the pattern of oscillation of neutrinos, causing the emergence of a damping factor in the neutrino transition probability. This might be significant in principle for long baseline-reactor neutrino experiments and for atmospheric neutrino observations. 
These are general results concerning all quantum spacetime-induced decoherence models that can be described in terms of a Lindblad-type  equation for the evolution of neutrinos. However, whether a specific model affects observations to a detectable level depends on the specific decoherence-inducing mechanism. These predict different forms of the damping factor, characterized by the dependence on the neutrino energy and mass. 
Among the models we considered, we found that only the one where decoherence is induced by stochastic metric fluctuations \cite{Breuer:2008rh} can be significantly constrained using reactor and atmospheric neutrinos. With these, we set conservative constraints on the scale governing the strength of the decoherence process that are several orders of magnitude stronger than the Planck scale, $E_{QG}\geq 2.6\cdot 10^{34}\;\text{GeV}\;$ and $E_{QG}\geq 2.5\cdot 10^{55}\;\text{GeV}\;$ from the KamLAND reactor neutrino experiment and the Super-Kamiokande atmospheric neutrino observatory, respectively. These  constraints would require the scale governing the model to be unrealistically  large, so that the hypothesis that the decoherence induced by stochastic metric perturbations affects neutrino oscillations  can be considered to be ruled out.

\section*{Acknowledgements}
We acknowledge financial support by the Programme STAR Plus, funded by Federico II University and Compagnia di San Paolo, and by the MIUR, PRIN 2017 grant 20179ZF5KS. This work contributes to the European Union COST Action CA18108 {\it Quantum gravity phenomenology in the multi-messenger approach.}

\newpage

\appendix

\section{Relativistic model with stochastic metric fluctuations}\label{appendix relativistic model}\renewcommand{\theequation}{\thesection.\arabic{equation}}\setcounter{equation}{0}

To study the effects of decoherence induced by stochastic metric perturbations on neutrino oscillations, in this appendix we derive a stochastic Schr\"{o}dinger equation resulting from the metric fluctuations defined in \cite{Breuer:2008rh}, adapted to a relativistic regime. 

Consider a particle evolving freely in a spacetime with stochastic perturbations. Denoting by $\tau$ the proper time of the particle and with $t$ the laboratory time, the Schr\"{o}dinger equation in the laboratory time reads \cite{Zych:2011hu}

\begin{equation}
    i\hbar \dv{}{t}\ket{\psi} = \dot{\tau}H\ket{\psi}\;,
\end{equation}
where $H$ is the Hamiltonian with respect to the proper time $\tau$ and

\begin{equation}
    \dot{\tau} = \frac{1}{c}\sqrt{-g_{\mu\nu}\dv{x^\mu}{t}\dv{x^\nu}{t}}\;.
\end{equation}
Writing the metric as $g_{\mu\nu} = \eta_{\mu\nu}+h_{\mu\nu}$, with $\abs{h_{\mu\nu}}\ll \abs{\eta_{\mu\nu}}$ being a (time-dependent) stochastic perturbation, we get

\begin{equation}
    \dot{\tau} = \gamma^{-1}\sqrt{1-\frac{h_{\mu\nu}p^\mu p^\nu}{m^2 c^2}} \sim \gamma^{-1}\qty(1-\frac{h_{\mu\nu}p^\mu p^\nu}{2m^2 c^2})\;,
\end{equation}
where $\gamma$ is the Lorentz factor and the last expansion comes from $\eta_{\mu\nu}p^\mu p^\nu = -m^2 c^2$ and $\abs{h_{\mu\nu}}\ll \abs{\eta_{\mu\nu}}$. By calling $H_0 = \gamma^{-1}H$ the free Hamiltonian in the laboratory frame and $H_s = -\frac{h_{\mu\nu}p^\mu p^\nu}{2m^2 c^2}H_0$ the stochastic correction, we get a stochastic Schr\"{o}dinger equation 

\begin{equation}
     i\hbar \dv{}{t}\ket{\psi} = \qty(H_0+H_s)\ket{\psi}\;.
\end{equation}
By following the same techniques in \cite{Petruzziello:2020wkd} (namely, by means of a cumulant expansion \cite{VANKAMPEN1974215}) and assuming the following form for the correlator of the stochastic fluctuations \cite{Breuer:2008rh}

\begin{equation}
    \expval{h_{ij}(t)h_{mn}(t^\prime)} = \tau_c \delta_{ij}\delta_{mn}\delta(t-t^\prime)\;,
\end{equation}
with all the other correlators being $0$ and with $\tau_c$ being the characteristic time scale of the perturbations, we get the Lindblad equation

\begin{equation}
    \partial_t \rho = -\frac{i}{\hbar}\comm{H_0}{\rho} - \frac{\tau_c}{\hbar^2}\comm{\frac{\pmb{p}^2}{2m^2c^2}H_0}{\comm{\frac{\pmb{p}^2}{2m^2c^2}H_0}{\rho}}\,.
\end{equation}
By restoring natural units ($\hbar=c=1$), and by setting the scale of the fluctuations to be the quantum gravity scale $\tau_c= E_{QG}^{-1}$, we see that the Lindblad operator defining this equation is given by

\begin{equation}\label{lindblad relativistic model}
    \mathcal{L} = \frac{\pmb{p}^2}{2m^2}\frac{H_0}{\sqrt{E_{QG}}}\;,
\end{equation}
which leads to a damping factor

\begin{equation}
    D_{ij} = \frac{L\,E_{ij}^6 \qty(\Delta m^2_{ij})^2}{4v_{g_ij}E_{QG}\,m_i^4m_j^4}\;.
\end{equation}

\newpage

\providecommand{\href}[2]{#2}\begingroup\raggedright\endgroup

\end{document}